\documentclass[prb,twocolumn,amssymb,amsmath,aps,superscriptaddress,showpacs]{revtex4}
\usepackage{amsfonts}
\usepackage{graphicx}
\usepackage{epsfig}
\def\bea{\begin{eqnarray}}
\def\eea{\end{eqnarray}}
\def\be{\begin{equation}}
\def\ee{\end{equation}}

\usepackage{hyperref}

\begin{document}
\title{Nonequilibrium conductance of asymmetric nanodevices in the Kondo regime}

\author{Eran Sela and Justin Malecki\date{\today}}
\affiliation{Department of Physics and Astronomy, University of
British Columbia, Vancouver, B.C., Canada, V6T 1Z1}

\begin{abstract}
The scaling properties of the conductance of a Kondo impurity
connected to two leads that are in or out of equilibrium has been
extensively studied, both experimentally and theoretically.  From
these studies, a consensus has emerged regarding the analytic
expression of the scaling function. The question addressed in this
brief report concerns the properties of the experimentally
measurable coefficient $\alpha$ present in the term describing the
leading dependence of the conductance on $eV/T_K$, where $V$ is the
source-drain voltage and $T_K$ the Kondo temperature. We study the
dependence of $\alpha$ on the ratio of the lead-dot couplings for
the particle-hole symmetric Anderson model and find that this
dependence disappears in the strong coupling Kondo regime in which
the charge fluctuations of the impurity vanish.
\end{abstract}
\pacs{75.20.Hr, 71.10.Hf, 75.75.+a, 73.21.La}

\maketitle

Universality of scaling plays a central role in the Kondo
effect~\cite{Kondo64} which describes the interaction of a magnetic
impurity with conduction electrons. As the temperature $T$ is
lowered below a crossover scale, denoted by the Kondo temperature
$T_K$, the impurity spin becomes screened by conduction electrons.
Recent advances in nanofabrication techniques now allow for the
experimental exploration of Kondo physics by attaching two
conducting leads, which we denote left L and right R, to a smaller
region, such as a semiconductor quantum dot (QD) or a single
molecule transistor (SMT), each of which may act as an effective
impurity spin. One typically applies a voltage difference $V$
between the leads and measures the differential conductance $G(T,V)
= dI/dV$ which grows at low $T$ and $V$ due to Kondo correlations.
This behavior is expected to be described by a scaling form
$G(T,V)=G_0 F(T/T_K,eV/T_K)$ where $G_0 = G(0,0)$ and $F$ is a
universal scaling function. The scale $T_K$ can vary between $100
{\rm{mK}}$ in QD devices~\cite{Gordon98,Cronenwett98} and $150
{\rm{K}}$ in SMTs.~\cite{Park02,Liang02}

Such devices are more accurately described by the Anderson model
(AM) which takes into account charge fluctuations of the impurity.
The Kondo model (KM) is recovered from the particle-hole symmetric
AM in the limit $U \to \infty$ where $U$ is the charging energy in
the AM. For both models, for low energies $T,eV \ll T_K$ the leading
corrections to the conductance are given by (up to a redefinition of
$T_K$)
\begin{equation}
\label{eq:alpha} F(T/T_K,eV/T_K) = 1-c_T
\bigl(\frac{T}{T_K}\bigr)^2-\alpha c_T  \bigl(
\frac{eV}{T_K}\bigr)^2 + ...~,
\end{equation}
with various values predicted for the coefficient $\alpha$ from the
AM~\cite{Costi94,Konik02,Oguri05B,Rincon09} and Kondo model
KM.~\cite{Schiller95,Majumdar98,Pustilnik04,Kaminski00} Recently,
$\alpha$ was measured by two experiments, one done by Grobis
\emph{et. al.} on a quantum dot device~\cite{Grobis08} and the other
done by Scott \emph{et. al.} on an ensemble of single molecule
transistors,~\cite{Scott09} where the QD and SMTs were tuned to the
Kondo regime $T,eV \ll T_K$. In the QD experiment, $T_K$ varied from
$150$ to $300{\rm{mK}}$ by varying the gate voltage in a single
device, and a value $\alpha_{{\rm{QD}}} = 0.1 \pm 0.015$ was
measured. In the SMTs experiment, $T_K$ ranged from $34$ to
$155{\rm{K}}$ in $29$ different devices, and $\alpha$ showed a
systematic deviation from the QD value, $\alpha_{{\rm{SMT}}} = 0.051
\pm 0.01$ (see Refs.~\onlinecite{Grobis08,Scott09} for the precise
fitting range). Various possible explanations were pointed out for
the systematic difference of $\alpha$ in SMTs. Among those, the
relative asymmetry of the L and R coupling [denoted by $A$ in
Eq.~(\ref{eq:A})] was considered as a relevant issue.~\cite{Scott09}

With this experimental motivation, we calculate $\alpha$ for
arbitrary device asymmetry. We consider the particle-hole symmetric
Anderson model (SAM) which generically includes charge fluctuations
in the impurity. We find that $\alpha$ is independent of the degree
of L-R asymmetry only in the Kondo limit. Once charge fluctuations
are included, there is a dependence of $\alpha$ on the L-R
asymmetry. We compare our result to previous theoretical results and
also comment on the relevancy to the experiments. We find that the
low value of $\alpha$ measured in SMTs can not be accounted for
within the symmetric Anderson model.

Our phenomenological approach consists of a modification of
Nozi\'{e}res Fermi liquid (FL) theory~\cite{Nozieres74} to account
for charge fluctuations in the SAM. We pay special attention to the
effect of shifting the Kondo resonance at finite voltage [see
Eq.~(\ref{eq:muK})]. Our result is a generalization of Oguri's
calculation of the conductance which used a non-perturbative result
for the Green's function of the SAM,~\cite{Oguri05B} and is found to
reduce to Oguri's result for the special case he considered.

The model of a single Anderson impurity connected to L and R leads
is
\begin{eqnarray}
H = H_{0}+H_d+H_t, \nonumber \\
H_0 = \sum_{k \sigma} \sum_{i=L,R} \epsilon_{k} c^\dagger_{k \sigma
i} c_{k \sigma i},
\end{eqnarray}
\begin{eqnarray}
H_d=\epsilon_d \sum_\sigma d^\dagger_\sigma d_\sigma+ U
d^\dagger_\uparrow d_\uparrow d^\dagger_\downarrow
d_\downarrow,\nonumber \\
 H_t =\sum_{k \sigma}  \sum_{i=L,R} v_i
\left(d^\dagger_\sigma c_{k \sigma i}+{\rm{ H.c.}}\right),
\end{eqnarray}
where $d_\sigma$ annihilates an electron with spin $\sigma$ in the
quantum dot $d$-level, $c_{k \sigma i}$ annihilates a conduction
electron with momentum $k$ and spin $\sigma$ in the $i=L,R$ lead,
$\epsilon_k = \hbar v_F k$, and $v_F$ is the Fermi velocity. In the
SAM that we will consider here, $\epsilon_d=-U/2$. It is convenient
to define the L-R asymmetry parameter
\begin{eqnarray}\label{eq:A} {\rm{L-R~ asymmetry}}:~~~ A=\frac{4
\Gamma_L \Gamma_R}{(\Gamma_L+\Gamma_R)^2}.
\end{eqnarray} The chemical
potentials $\mu_i$ $(i=L,R)$, satisfying $\mu_L - \mu_R = eV$, are
measured relative the the Fermi level defined at zero voltage. Then
the ratio
\begin{equation}
B =  - \mu_L / \mu_R,
\end{equation}
describes the relative voltage drop across the L and R tunnel
junctions which could depend on the capacitive couplings of the
leads and QD or SMT and which we treat as a second L-R asymmetry
parameter.

One can define the retarded $d$ Green's function as $G^R(\epsilon)
=-i \int_{0}^\infty dt e^{i \epsilon t}  \langle d(t) d^\dagger +
d^\dagger d(t) \rangle$. With these definitions, the current is
given by  the Meir-Wingreen formula~\cite{Meir92}
\begin{equation}
\label{eq:Meir}I = \frac{2 e}{h} A \int_{- \infty}^\infty d
\epsilon[f_L-f_R] \Delta \left[- {\rm{Im}}G^R(\epsilon) \right],
\end{equation}
with $f_i = f(\epsilon-\mu_i)$ ($i=L,R$), $f(\epsilon)=[1+\exp(
\epsilon/T)]^{-1}$, $\Delta =\Gamma_L+\Gamma_R $, $\Gamma_i = \pi
\nu v_i^2$ ($i=L,R$), and $\nu$ is the density of states in the
leads.

Below, we formulate an effective theory for the SAM that allows us
to obtain the Green's function $G^R(\epsilon)$ for $\epsilon, T, eV
\ll \tilde{\Delta}$ [$\tilde{\Delta}$ is the characteristic energy
scale defined more explicitly below Eq.~(\ref{eq:phaseshift}). In
the Kondo limit $\tilde{\Delta} \to T_K$]. We define linear
combinations of the annihilation operators for the L and R leads
\begin{equation}
\label{eq:sp} a_{k \sigma}^{(s)} = \frac{v_L c_{k \sigma L}+v_R c_{k
\sigma R}}{v}, ~~~a_{k \sigma}^{(p)} =\frac{- v_R c_{k \sigma L}+v_L
c_{k \sigma R}}{v},
\end{equation}
where $v=\sqrt{|v_L|^2+|v_R|^2}$. Only the $s-$wave particles are
coupled to the $d$-level, since $H_t = \sum_{k \sigma} v
 d^\dagger_\sigma a^{(s)}_{k \sigma}+{\rm{ H.c.}}$.

The notion of a local Fermi liquid (FL), due to
Nozi\'{e}res,~\cite{Nozieres74} was originally applied to the Kondo
model but is actually more general and can be applied to the AM and,
in particular, to the less complicated SAM. The quasiparticles of
this FL theory are simply scattering states whose incoming part
coincides with that of the the $s-$wave particles $a_{k
\sigma}^{(s)}$ [the precise definition is given after
Eq.~(\ref{eq:fp})]. The theory itself consists of a low energy
expansion of their scattering phase shift as a function of energy
$\epsilon$ (measured from the Fermi level) and of quasiparticle
density $n_\sigma$,
\begin{equation}
\label{eq:phaseshift} \delta_{\sigma} = \delta_0+
\frac{\epsilon}{\tilde{\Delta}}-\frac{\beta n_{\bar{\sigma}}}{\nu
\tilde{\Delta}}+...~,
\end{equation}
where $\bar{\uparrow} = \downarrow$, $\bar{\downarrow} = \uparrow$,
$\tilde{\Delta}$ is the energy scale over which the phase shift
varies in the SAM, and $\beta$ is a coefficient to be determined.
For the general AM, the scattering phase shift at the Fermi energy
$\delta_0$ can be extracted using the Friedel sum
rule~\cite{Langreth66} $\delta_0 = {\rm{Im}} [\ln G^R(0)|_{T=eV=0}]
- \pi$ combined with exact results for $G^R$. In the SAM,
particle-hole symmetry and the adiabatic connection to the $U=0$
case implies $\delta_0=\pi/2$.

The Wilson ratio $R=(\delta \chi / \chi)/(\delta C_v/C_v)$ is the
ratio between the relative impurity contribution to the
susceptibility and to the specific heat. It can be calculated from
the phase shift expansion, Eq.~(\ref{eq:phaseshift}), to
be~\cite{Nozieres74}
\begin{equation}
\label{eq:beta} R=1+\beta.
\end{equation}
We will use this equation to determine $\beta$ in terms of $R$ which
is a parameter describing the amount of charge fluctuations in the
SAM.

As an equivalent way to determine $\beta$, consider an enhancement
of the Fermi energy by an amount $\epsilon$ by adding to the Fermi
sea a density of quasiparticles $n_\sigma=(\nu \epsilon)$. For the
KM, Nozi\'{e}res argued that at energy $\epsilon$, corresponding to
the new Fermi energy, one has $\delta_{\sigma} = \delta_0$ since the
Kondo resonance is tied to the Fermi level. Using
Eq.~(\ref{eq:phaseshift}) and $n_{\bar{\sigma}} = \nu \epsilon$,
this implies $\beta=1$. This argument should be modified for the AM:
a shift of the Fermi level by this transformation also implies that
$\epsilon_d \to \epsilon_d - \epsilon$, as measured relative to the
new Fermi level at $+ \epsilon$. Therefore a finite amount of charge
$e \delta n_d$ enters into the $d$-level which, for small
$\epsilon$, is given in terms of the charge susceptibility $\delta
n_d = -\frac{dn_d}{d \epsilon_d} \epsilon$. Here $n_d=\sum_\sigma
\langle d^\dagger_\sigma d_\sigma \rangle$. Due to the Friedel sum
rule,~\cite{Langreth66} the phase shift at the new Fermi energy
after this transformation is different than $\delta_0$ and given by
$\delta_\sigma = \pi n_d/2 = \delta_0 -(\pi/2) \frac{dn_d}{d
\epsilon_d} \epsilon$. Using Eq.~(\ref{eq:phaseshift}) and
$n_{\bar{\sigma}} = \nu \epsilon$ implies $\beta=1+\frac{\pi
\tilde{\Delta}}{2} \frac{dn_d}{d \epsilon_d}$.

This phase shift expansion can be equivalently described by a
Hamiltonian
\begin{eqnarray}
\label{eq:fp} H &=& H_0(a^{(p)})+H_0(b)+\delta H,\nonumber \\ \delta
H &=& -\frac{1}{2 \pi \nu \tilde{\Delta}} \sum_{k k'
\sigma}(\epsilon_k+\epsilon_{k'}) b^\dagger_{k \sigma} b_{k'
\sigma}\nonumber
\\&&+\frac{R-1}{ \pi \nu^2 \tilde{\Delta}} \sum_{k_1 k_2 k_3 k_4}:
b^\dagger_{k_1 \uparrow} b_{k_2 \uparrow} b^\dagger_{k_3 \downarrow}
b_{k_4 \downarrow}:,
\end{eqnarray}
where $H_0 (\psi)=\sum_{k \sigma} \epsilon_k \psi_{k \sigma}^\dagger
\psi_{k \sigma} $ ($\psi=a^{(p)},b$). This Hamiltonian describes the
two last terms in the phase shift expansion
Eq.~(\ref{eq:phaseshift}).

The first term in Eq.~(\ref{eq:phaseshift}), $\delta_0$, is
incorporated into the definition of the $b$-particles in
Eq.~(\ref{eq:fp}).  These $b$-particles are single particle
scattering states that describe an incoming $s$-wave suffering a
scattering phase shift $\delta_0$ at the boundary. Formally, to
define the $b$-particles, one uses the unfolding
transformation~\cite{Affleck91} where $\psi^{(s)}_\sigma (x)= (1/2
\pi) \int dk e^{-i k x} a_{k \sigma}^{(s)}$ ($x \in \{ -
\infty,\infty \}$) is a chiral field describing an $s$-wave
scattering state with the left moving convention such that $x>0$ is
the incoming part and $x<0$ the outgoing part, $x=0$ being the
boundary. From the definition of the phase shift $\delta_0$ we have
$\psi^{(s)}_\sigma (0^-) = e^{2 i \delta_0} \psi^{(s)}_\sigma (0^+)
$. We define the $b-$particles in terms of a scattering state with
$\delta_0 = \pi / 2$
\begin{eqnarray}
\label{eq:transformation} \psi^{(b)}_\sigma (x)=\psi^{(s)}_\sigma
(x){\rm{sgn}}(x),
\end{eqnarray}
and its Fourier modes $b_{k \sigma} = \int dx e^{i k x}
\psi^{(b)}_\sigma (x)$.

Now consider $eV \ne 0$. As long as $eV \ll \tilde{\Delta}$ the
system remains in the vicinity of the fixed point and the state at
finite $eV$ can be treated within the FL theory as a state with a
non-thermal distribution of quasiparticles. We first consider single
particle scattering states incoming from lead $i=L,R$. In second
quantization those particles are annihilated by $(c_{k \sigma
i})^{in}$. The occupation of those incoming waves is simply $\langle
({c^\dagger_{k \sigma i}})^{in} ({c_{k' \sigma' i'}})^{in} \rangle =
\delta_{k k'} \delta_{\sigma \sigma'} \delta_{i i'}
f_i(\epsilon_k)$. Using Eq.~(\ref{eq:sp}), and the fact that before
the scattering region ($x>0$) the wave function of states
$a_{k\sigma}^{(s)}$ and $b_{k\sigma}$ coincide, we
have~\cite{Kaminski00}
\begin{eqnarray}
(c_{k \sigma L })^{in}=(v_L b_{k \sigma} - v_R a^{(p)}_{k \sigma})/v, \nonumber \\
(c_{k \sigma R} )^{in}=( v_R b_{k \sigma} + v_L a^{(p)}_{k
\sigma})/v. \nonumber
\end{eqnarray}
This gives the nonequilibrium distribution function for the
$b$-particles
\begin{equation}
\label{eq:noneqdis} \langle b^\dagger_{k \sigma } b_{k' \sigma' }
\rangle = \delta_{k k'} \delta_{\sigma \sigma'} \left[ \Gamma_L
f_L(\epsilon_k)+\Gamma_R f_R(\epsilon_k) \right]/\Delta.
\end{equation}

Since the occupation of the $b$-particles differs from the one
defined at $T=eV=0$, the second term of $\delta H$ generates a
constant elastic scattering $\frac{ (R-1) n_{\bar{\sigma}} }{ \pi
\nu^2 \tilde{\Delta}} \sum_{k_1 k_2 \sigma} b^\dagger_{k_1 \sigma}
b_{k_2 \sigma}$, where
\begin{eqnarray}
\label{eq:alphab}
 n_{\bar{\sigma}} &=&\sum_k \langle :b^\dagger_{k \bar{\sigma}} b_{k \bar{\sigma}}:\rangle  =\nu (\Gamma_L \mu_L+\Gamma_R
 \mu_R)/\Delta.
\end{eqnarray}
As a result, the phase shift at energy $\epsilon$ relative to the
Fermi energy is given by [see Eq.~(\ref{eq:phaseshift})]
\begin{equation}
\label{eq:deltas} \delta_s(\epsilon)=\frac{\pi}{2}+\frac{\epsilon -
\mu_K}{\tilde{\Delta}},
\end{equation} where $\mu_K$ is a \emph{nonequilibrium shift of the resonance},
\begin{equation}
\label{eq:muK} \mu_K = (R-1) (\Gamma_L \mu_L+\Gamma_R \mu_R)/\Delta.
\end{equation}
This shift can be nonvanishing if $e V \ne 0$, and $U>0$ (or
equivalently $R \ne 1$). In the Kondo limit, $R=2$,
Eq.~(\ref{eq:muK}) implies that the resonance position $\mu_K$
shifts together with the average chemical potential $\bar{\mu} =
(\mu_L+\mu_R)/2$: under $\bar{\mu} \to \bar{\mu} + \delta \mu$ at
fixed $eV$, Eq.~(\ref{eq:muK}) gives $\mu_K \to \mu_K + \delta \mu$.

In order to calculate the current using the phenomenological
Hamiltonian Eq.~(\ref{eq:fp}), one can use the Meir-Wingreen
formula, and relate the Green's function $G^R(\epsilon)$ to the
$s$-wave $T$-matrix, $T_s(\epsilon) = v G^R(\epsilon) v$. One can
define a $T-$matrix, $\tilde{T}$, for the $b$-particles due to
$\delta H$. It has an inelastic part denoted by $\tilde{T}_{in}$.
The relation between $T_s$ and $\tilde{T}$, accounting for the small
inelastic term $\tilde{T}_{in}$, reads~\cite{Pustilnik04}
\begin{equation}
\label{eq:pustGla} -\pi \nu T_s(\epsilon) = \frac{1}{2i}  \left[
e^{2 i \delta_s(\epsilon)}-1 \right]+ e^{2 i \delta_s(\epsilon)}
\left[- \pi \nu  \tilde{T}_{in}(\epsilon) \right].
\end{equation}

\begin{figure}[h]
\begin{center}
\includegraphics*[width=15mm]{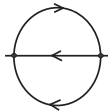}
\caption{Diagram used to calculate $\tilde{T}_{in}$. Each line is a
$b$-particle propagator, with the nonequilibrium distribution
Eq.~(\ref{eq:noneqdis}). \label{fg:2}}
\end{center}
\end{figure}

The leading contribution to $\tilde{T}_{in}$ originates from the
diagram shown in Fig.~\ref{fg:2}, containing three propagators of
$b$-particles whose occupation is given by Eq.~(\ref{eq:noneqdis}).
We find, using the Keldysh technique,\widetext
\begin{eqnarray}
\label{eq:Tin} - \pi \nu {\rm{Im}}
\tilde{T}_{in}(\epsilon)&=&\frac{(R-1)^2}{4 \tilde{\Delta}^2}\int d
\epsilon_{1} d \epsilon_{2} d \epsilon_{3}[1+t(\epsilon_2)
t(\epsilon_3)-t(\epsilon_1)\{t(\epsilon_2)+ t(\epsilon_3)\}]
\delta(\epsilon+\epsilon_{1}-\epsilon_{2}-\epsilon_{3})
\nonumber \\
&=&\frac{(R-1)^2}{2 \tilde{\Delta}^2} [ \pi^2 T^2 +(\epsilon -\kappa
eV )^2+ (eV)^2 \frac{3A}{4}],
\end{eqnarray}
where $t(\epsilon) =1-2 \left[ \Gamma_L f_L(\epsilon)+\Gamma_R
f_R(\epsilon) \right]/\Delta$ and $\kappa = (\Gamma_L \mu_L+\Gamma_R
\mu_R)/[\Delta (\mu_L - \mu_R)] = (B \Gamma_L - \Gamma_R) / [(1+B)
\Delta]$. Plugging Eqs.~(\ref{eq:deltas}) and (\ref{eq:Tin}) into
Eq.~(\ref{eq:pustGla}) and using the Meir-Wingreen formula
Eq.~(\ref{eq:Meir}), with $G^R(\epsilon)=v^{-2} T_s(\epsilon)$, we
obtain the conductance in the scaling form of Eq.~(\ref{eq:alpha})
with $G_0=\frac{2 e^2}{h} A$, $c_T = \frac{\pi^2 [1+2(R-1)^2]}{3}$,
$T_K \to \tilde{\Delta}$ and
\begin{eqnarray}
\label{eq:alphaLR} \alpha = \frac{9}{2 \pi^2} \left( \kappa
(R^2-1)\frac{1-B}{1+B}+\frac{2+(R-1)^2}{3} \frac{1+B^3}{(1+B)^3}
+3(R-1)^2 (\kappa^2+\frac{1}{4}A)\right) / \left(1+2(R-1)^2 \right).
\end{eqnarray}
\endwidetext The coefficient $\alpha$ can be expressed as a function of 3 independent variables such as the Wilson ratio $R$
and the L-R asymmetry parameters $A$ and $B$. In general, $\alpha$
depends on the the L-R asymmetry parameters however, in the strong
coupling limit $U \to \infty$, equivalent to $R \to 2$, this
dependence completely disappears from Eq.~(\ref{eq:alphaLR}) and we
obtain $\alpha[R=2] = \frac{3}{2 \pi^2}=0.1519$. In this limit,
$\tilde{\Delta} \to 4 \pi \sqrt{u/2 \pi} \exp[-\pi^2 u/8+1/(2u)]$
which is the known expression for $T_K$, where $u=U/(\pi \Delta)$.

The value of $\alpha$ measured in QDs can be accounted for by charge
fluctuations due to finite $U$ since, for $R \ne 2$,
Eq.~(\ref{eq:alphaLR}) gives $\alpha[R ,A,B]$ in the range $3/4
\pi^2=0.0759 \le \alpha \le 0.3039= 3/\pi^2$.  Since one can tune
the gate voltage in QDs and this corresponds to tuning $\epsilon_d$,
it is plausible that the SAM applies at one value of the gate
voltage corresponding to maximal conductance. However the value
$\alpha_{{\rm{SMT}}} = 0.051 \pm 0.01$ measured in SMTs is lower
than the expectation from the SAM. We note that a value $\alpha =
0.157 \pm 0.005$ was also measured~\cite{Lin} in $\rm{Al/AlO_x/Sc}$
planar tunnel junctions at low temperatures~\cite{Yeh09} and can be
accounted for in our theory.

We compare Eq.~(\ref{eq:alphaLR}) to the results of other approaches
for the KM and SAM. Firstly, our result is fully consistent with the
results of Oguri~\cite{Oguri05B} based on Ward identities, however
he concentrated on the special case of $A=B=1$; Kaminski, Nazarov,
and Glazman~\cite{Kaminski00} find $\alpha_{{\rm{KNG}}} = \frac{3}{8
\pi^2}$ for the KM for any $A$; Konik, Saleur, and
Ludwig~\cite{Konik02} find $\alpha_{{\rm{KSL}}}=4/\pi^2$ for the SAM
for $A=1$ and large $U/\Delta$; Pustilnik and
Glazman~\cite{Pustilnik04} find for the KM for $A \ll 1$,
$\alpha_{{\rm{PG}}} = \frac{3}{2 \pi^2}$; Rinc\'{o}n, Aligia, and
Hallberg~\cite{Rincon09} studied the SAM for the case $B =
\Gamma_R/\Gamma_L$. A mistake was found in Eq.~(10) in their paper,
whereas the corrected formula~\cite{remark} $
 \frac{G}{G_{0}}|_{B = \frac{\Gamma_R}{\Gamma_L}} \simeq 1-\frac{\pi ^{2}(1+2(R-1)^{2})}{3}\left(
 \frac{T}{\tilde{\Delta}}\right) ^{2}  -\frac{4-3A+(2+3A)(R-1)^{2}}{4}\left(
 \frac{eV}{\tilde{\Delta}}
 \right) ^{2}$ agrees with our
Eq.~(\ref{eq:alphaLR}). Given that $\alpha_{{\rm{KSL}}}$, which
differs from our result, was obtained approximately and is not
claimed to be exact, and given the agreement with the exact
formulation due to Oguri,~\cite{Oguri05B} we are convinced of the
validity of our reported expression for $\alpha$.

As another application of Eq.~(\ref{eq:fp}), one can calculate the
shot noise $S = \int_{- \infty}^\infty dt \langle  \{ \delta I(t),
\delta I \} \rangle$ for the SAM to leading order in
$1/\tilde{\Delta}$ where $\delta I = I - \langle I \rangle$. At
$T=0$ and $A=B=1$, using results for $S$ based on the effective
Hamiltonian $\delta H$ of Eq.~(\ref{eq:fp}) with arbitrary
coefficients,~\cite{Gogolin06,Sela06Frac} we obtain $S = \frac{4
e^2}{h}\frac{1+9(R-1)^2}{12}\left(\frac{eV}{\tilde{\Delta}}
\right)^2$. The ratio 
\begin{equation}
e^* \equiv \frac{S}{2 I_b}=\frac{1+9
(R-1)^2}{1+5 (R-1)^2} e ,
\end{equation}
where $I_b=\frac{2e^2}{h}V - I$, can be
interpreted as a backscattering charge.~\cite{Sela06Frac}  This
charge crosses from $e^*=e$ for $R=1$ (noninteracting resonance
level) to $e^*=5/3 e$ for $R=2$ (Kondo
resonance).~\cite{Sela06Frac,Gogolin06,Fujii}

In conclusion, we extended Nozi\'{e}res Fermi liquid
theory~\cite{Nozieres74} to account for charge fluctuations in the
particle hole symmetric Anderson model and calculated the transport
coefficient $\alpha$ present in the term describing the leading
dependence on $eV/T_K$. We included explicitly the effects of L-R
asymmetry of the device and discussed the relation to recent
experimental results.~\cite{Grobis08,Scott09}

After this work was essentially completed, we became aware of
another work~\cite{SUN} that obtains $\alpha = \frac{3}{2 \pi^2}$
for the KM with arbitrary L-R asymmetry, consistent with our result
in the special case without charge fluctuations.

We thank I. Affleck, L. I. Glazman, D. Goldhaber-Gordon, Y. Oreg and
A. Rosch for fruitful discussions. This work was supported by NSERC.

\end{document}